\documentclass[usenatbib,useAMS,refree]{mnras}
\usepackage[T1]{fontenc}
\usepackage{ae,aecompl}
\usepackage{graphicx}
\usepackage{amsmath}
\usepackage{amssymb}
\newcommand\pulsar{2S~1417--624~}

\title[{\it NuSTAR} observation of 2S~1417-624]
{{\em NuSTAR} view of Be/X-ray binary pulsar 2S~1417--624 during 2018 giant outburst}

\author[Gupta et al.] 
{Shivangi Gupta$^1$\thanks{E-mail: shivangi@prl.res.in}, Sachindra Naik$^1$\thanks{snaik@prl.res.in}, 
Gaurava K. Jaisawal$^2$\thanks{gaurava@space.dtu.dk} \\
$^1$ Astronomy and Astrophysics Division, Physical Research Laboratory, Navrangapura, Ahmedabad - 380009, Gujarat, India \\
$^2$ National Space Institute, Technical University of Denmark, Elektrovej 327-328, DK-2800 Lyngby, Denmark\\  }

\begin{document}

\date{}

\maketitle

\begin{abstract}

We report the results obtained from a detailed timing and spectral studies of Be/X-ray binary pulsar \pulsar using data from {\it Swift} and {\it NuSTAR} observatories. The observations were carried out at the peak of a giant outburst of the pulsar in 2018. X-ray pulsations at  17.475~s were detected in the source light curves up to $\sim$79~keV. The evolution of the pulse profiles with energy was found to be complex. A four-peaked profile at lower energies gradually evolved into a double-peak structure at higher energies. The pulsed fraction of the pulsar, calculated from the {\it NuSTAR} observation was found to follow an anti-correlation trend with luminosity as observed during previous giant X-ray outburst studies in 2009. The broadband spectrum of the pulsar is well described by a composite model consisting of a cutoff power law model modified with the interstellar absorption, a thermal blackbody component with a temperature of $\approx$1~keV and a Gaussian function for the 6.4 keV iron emission line. Though the pulsar was observed at the peak of the giant outburst, there was no signature of presence of any cyclotron line feature in the spectrum. The radius of the blackbody emitting region was estimated to be $\approx$2~km, suggesting that the most probable site of its origin is the stellar surface of the neutron star. Physical models were also explored to understand the emission geometry of the pulsar and are discussed in the paper.

\end{abstract}


\section{Introduction}

Be/X-ray binary pulsars are known to exhibit two types of X-ray outbursts, such as Type~I and Type-II X-ray outbursts \citep{PaulNaik2011, Reig2011}. Type-I outbursts are periodic X-ray events that usually occur at the periastron passage of the neutron star in the Be/X-ray binary systems. These events last for $\sim$20--30\% duration of the binary orbit during which the peak X-ray luminosity of the pulsar reaches as high as $\sim$10$^{37}$ erg s$^{-1}$. However, there are X-ray outbursts observed from these Be/X-ray binary pulsars during which the peak luminosity exceeds 10$^{37}$~erg~s$^{-1}$. These events are known as Type-II X-ray outbursts. These events are observed occasionally for a duration of a few weeks to a few months and are independent of the orbital phase of the binary. Irrespective of the outburst type, the observed X-ray radiation from the pulsars originates from hot spots at the magnetic poles of the pulsar and/or from column-like structures mounted on the surface of the magnetized neutron stars. These column-like structures are called accretion columns which are formed due to the channelling of accreted matter by the magnetic field of the neutron stars. 

During the process of accretion, matter gains sufficient amount of kinetic energy as it approaches the magnetic poles of the neutron star. Corresponding amount of energy is released in the form of thermal radiation when the matter hits the poles of the neutron star \citep{Basko1976, Becker2012}. The process of interaction and geometry of the emitting regions at the magnetic poles of the neutron star depends on the mass accretion rate (luminosity) of the pulsar. At low mass accretion rate, the hot spots at the magnetic poles of the neutron star is the source of X-ray emission. At higher mass accretion rate, a radiation dominated shock front appears in the accretion column that shapes the X-ray emission  through bulk and thermal Comptonization processes from these sources. Depending on mass accretion rate, a  transition between sub-critical and super-critical regimes \citep{Basko1976, Becker2012, Mushtukov2015} is also known in several X-ray pulsars such as V~0332+53 \citep{Doroshenko2017}, EXO~2030+375 \citep{Epili2017}, Swift~J0243.6+6124 \citep{Wilson2018}. The transition between above two regimes is observed in terms of changes in the shape of pulse profiles (emission geometry) of the pulsar and its spectral index.

The transient Be/X-ray binary pulsar \pulsar was discovered with the {\it third Small Astronomy Satellite (SAS-3)} observations in 1978  \citep{Apparao1980}. It was successively detected with other space missions and recognized as 4U~1416--62 and MX~1418--61 (\citealt{Forman1978,Markert1979}). The analysis of {\it SAS-3} archival data revealed a coherent pulsation of $\sim$17.64~sec from the pulsar \citep{Kelley1981}. Later in 1979, observations with  High Resolution Imaging detector(HRI) detector onboard {\it Einstein} X-ray observatory provided the precise estimate of the source position. The optical companion was then identified to be a Be star located at a distance of 1.4--11.1~kpc \citep{Grindlay1984}. Like other Be/X-ray binary pulsars, \pulsar also exhibits Type~I and Type~II X-ray outbursts. Five Type~II and a number of Type~I outbursts have been detected from \pulsar so far (\citealt{Kelley1981, Finger1996, Inam2004, Gupta2018, Nakajima2018, Krimm2018}). The source has also been monitored during its quiescence phase in May 2013 \citep{Tsygankov2017a}.

Following its discovery in 1978, \pulsar remained in quiescence for about 16 years until a second giant outburst occurred in August 1994. The giant outburst lasted for about 110 days and subsequently followed by five smaller Type~I outbursts until July 1995 \citep{Finger1996}. BATSE onboard the Compton Gamma Ray Observatory ({\it CGRO}) continuously monitored the source during this period and the binary orbital parameters of the system were determined. The orbital period and eccentricity of the binary was found to be 42.12~d and 0.446, respectively \citep{Finger1996}. The mass function of the system was also derived from the BATSE data which put a lower limit of 5.9 M$_{\odot}$ on the mass of optical companion. The binary orbital parameters of the system were later refined by \citet{Raichur2010}, using data obtained from the {\it RXTE} observatory during the third giant outburst of the pulsar in 1999. Using the same set of observations (spanning a duration between 1999 November and 2000 August), \citet{Inam2004} reported the intensity dependent pulse profiles and pulsed fraction of the pulsar. The pulsed fraction was found to be correlated with the source flux and the pulse profiles consisted of two peaks, separated by a phase difference of $\sim$0.5. Also, the pulsar was observed to spin-up significantly during the outburst, which was interpreted as a sign of disc accretion. 

The fourth giant outburst from the pulsar was observed in November 2009, reaching a peak intensity of $\sim$300 mCrab in 15--50~keV energy band of {\it Swift}/BAT \citep{Krimm2009}. The complete outburst was regularly monitored with the {\it RXTE}. Analysis of these observations revealed a peculiar evolution of the pulse profiles and pulsed fraction with the source luminosity. The pulsed fraction was found to be anti-correlated with the source flux and pulse profiles evolved from double- to triple-peaked structure with increase in the source luminosity. Also, the energy resolved pulse profiles were found to exhibit a complex evolution. These observed changes were attributed to the change in beam pattern of the pulsar from pencil beam to a mixture of pencil and fan beam geometry \citep{Gupta2018}.

%%-----------------------------------------------------------------------------------------%%
\begin{figure}
\centering
\includegraphics[height=3.2in, width=2.4in, angle=-90]{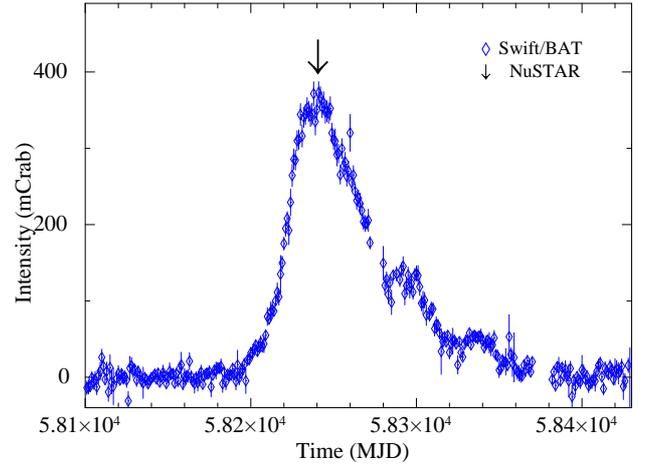}
\caption{{\it Swift}/BAT light curve of \pulsar in 15--50~keV energy range, covering a duration from 2017 December 18 (MJD 58105) to 2018 August 11 (MJD 58430). The arrow mark on the top of light curve indicates the start time of the {\it NuSTAR} observation (MJD 58240.57, see also Table~\ref{log}) during the 2018 giant outburst.}
\label{lightcurve_crab_norm}
\end{figure}
%%-----------------------------------------------------------------------------------------%%

The X-ray spectrum of the pulsar has often been modelled with an absorbed power law modified with a high energy cutoff (\citealt{Finger1996, Inam2004}), during its Type~I and Type~II outbursts. An additional iron line complex in $\sim$6.4--6.8~keV range was also detected during 1999 giant outburst \citep{Inam2004}. However, a cutoff power law model was found to describe the continuum equally well in later observations of \pulsar during the Type~II outburst in 2009 \citep{Gupta2018}. Recently, {\it Chandra} observed \pulsar in 2013 May during the quiescent phase. Pulsations were not detected in the source light curve, probably due to low exposure or low mass accretion rate \citep{Tsygankov2017a}. The pulsar spectrum was found to be ambiguous and described with either a power law or a blackbody model with a quite high temperature of $\sim$1.5~keV. The unexpected high blackbody temperature was interpreted as a result of continuing accretion from a cold disc (see \citealt{Tsygankov2017b} for details) depleted after 2009 outburst.

In the present work, we report timing and spectral properties of \pulsar using simultaneous observations with {\it NuSTAR} and {\it Swift} observatories during its recent 2018 May X-ray outburst (\citealt{Nakajima2018, Krimm2018}). It is the fifth and brightest giant outburst detected from the source till date. The paper is organized as follows: in Section~2, we describe the details of data reduction and analysis procedures, followed by timing and spectral analysis details in Section~3 and 4, respectively. In Section~5, we discuss our results in the context of previous outbursts and conclude with the concise summary of the work.

\section{Observations and Analysis}

Figure~\ref{lightcurve_crab_norm} shows the monitoring light curve of \pulsar during its 2018 giant X-ray outburst, as recorded by {\it Swift}/Burst Alert Telescope (BAT; \citealt{Krimm2013}). Arrow mark at the peak of the outburst (MJD 58240.57) indicates the beginning of the {\it NuSTAR} observation of the pulsar. {\it Swift}/XRT observed the same outburst for an effective exposure of $\sim$1.8~ks simultaneously (MJD 58240.81) with the {\it NuSTAR} observation. A log of all the observations used in the present work is listed in Table~\ref{log}. We have also used an {\it RXTE}/PCA \citep{Jahoda1996} light curve of \pulsar during the peak of 2009 outburst (ObsID 94032-02-04-01), for the pulse profile analysis which is presented in Section~3. Standard-1 binned mode data of PCA was used to extract the source light curve in 2--60~keV energy range with a time resolution of 0.125~s. After correcting for background contamination, barycentric correction was applied on the source light curve (see \citealt{Gupta2018} for the detailed reduction procedures). While the description of {\it NuSTAR} and {\it Swift} data reduction procedure is presented below. 

%%-----------------------------------------------------------------------------------------%%
\begin{table*}
\centering
\caption{Log of observations of \pulsar used in the present work.}
\begin{tabular}{ccccc}
\hline\\
Satellite/Instrument    &ObsID		&Start Time   &Exposure	&Pulse Period  \\
	&		&(MJD)		&(ksec)		&(sec)	\\
\hline
\\
{\em NuSTAR}/FPMs            &90402318002     &58240.57   &28.8,29.1$^1$ &17.475(6) \\
{\em  Swift}/XRT              &00088676001   	&58240.81   &1.8	   &17.47(5) \\
{\em RXTE}/PCA	       &94032-02-04-01	&55158.21   &3.4$^2$       &17.502(9) \\
\\
\hline
$^1$ : For FPMA and FPMB respectively.
$^2$ : See section 2,3 in text.
\end{tabular}
\label{log}
\end{table*}
%%-----------------------------------------------------------------------------------------%%

%-----------------------------------------------------------------------------------%
\begin{figure*}
\centering
\includegraphics[height=5.9in, width=3.6in, angle=-90]{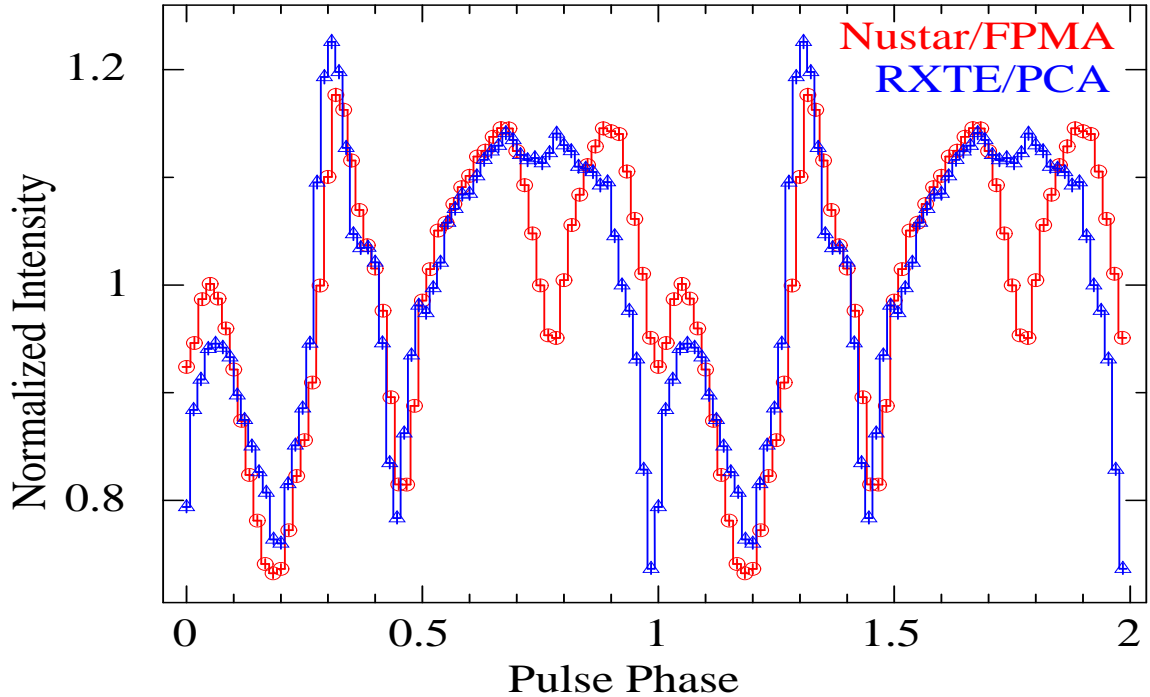}
\caption{The {\it NuSTAR} and {\it RXTE} pulse profiles of \pulsar at the peaks of 2018 and 2009 giant outbursts are shown in red circles and blue triangles, respectively. The pulse profiles are shown in 3--60~keV energy range for both the epochs of observations. Two pulses are shown in each panel for clarity. }
\label{fig2}
\end{figure*}
%-----------------------------------------------------------------------------------%

\subsection{{\it NuSTAR} observation}

{\it NuSTAR}, the first hard X-ray focusing observatory, is sensitive in 3--79~keV energy range \citep{Harrison2013}. It consists of two co-aligned identical grazing angle focusing telescopes with corresponding detector plane modules FPMA and FPMB. Each detector module comprises of four 32$\times$32 Cadmium-Zinc-Telluride (CZT) detectors arranged in a plane. The {\it NuSTAR} provides a spectral resolution of 0.4~keV at 10 keV and 0.9~keV at 68 keV (FWHM), respectively. 

The standard procedures were followed to analyze the {\it NuSTAR} data from each focal plane modules separately. The data were extracted using {\tt NUSTARDAS} pipeline (version 1.4.1), as distributed with {\tt HEASoft} version 6.16. The unfiltered events were first reprocessed by using {\tt NUPIPELINE} in the presence of updated version of Calibration data files (CALDB 20181030). Science quality events, obtained after reprocessing were utilized to extract the barycentric corrected light curves, spectra, effective area files and response matrices via {\tt NUPRODUCTS} package. A circular region of 120 arc-sec centered on the source coordinates was used to extract the source products from both the detectors (FPMA and FPMB) individually. The background light curves and spectra were accumulated by considering a circular region of the same size centered away from the source region in order to avoid contamination from the source photons. 

%-----------------------------------------------------------------------------------%
\begin{figure}
\centering
\includegraphics[width=4.4in, angle=-90]{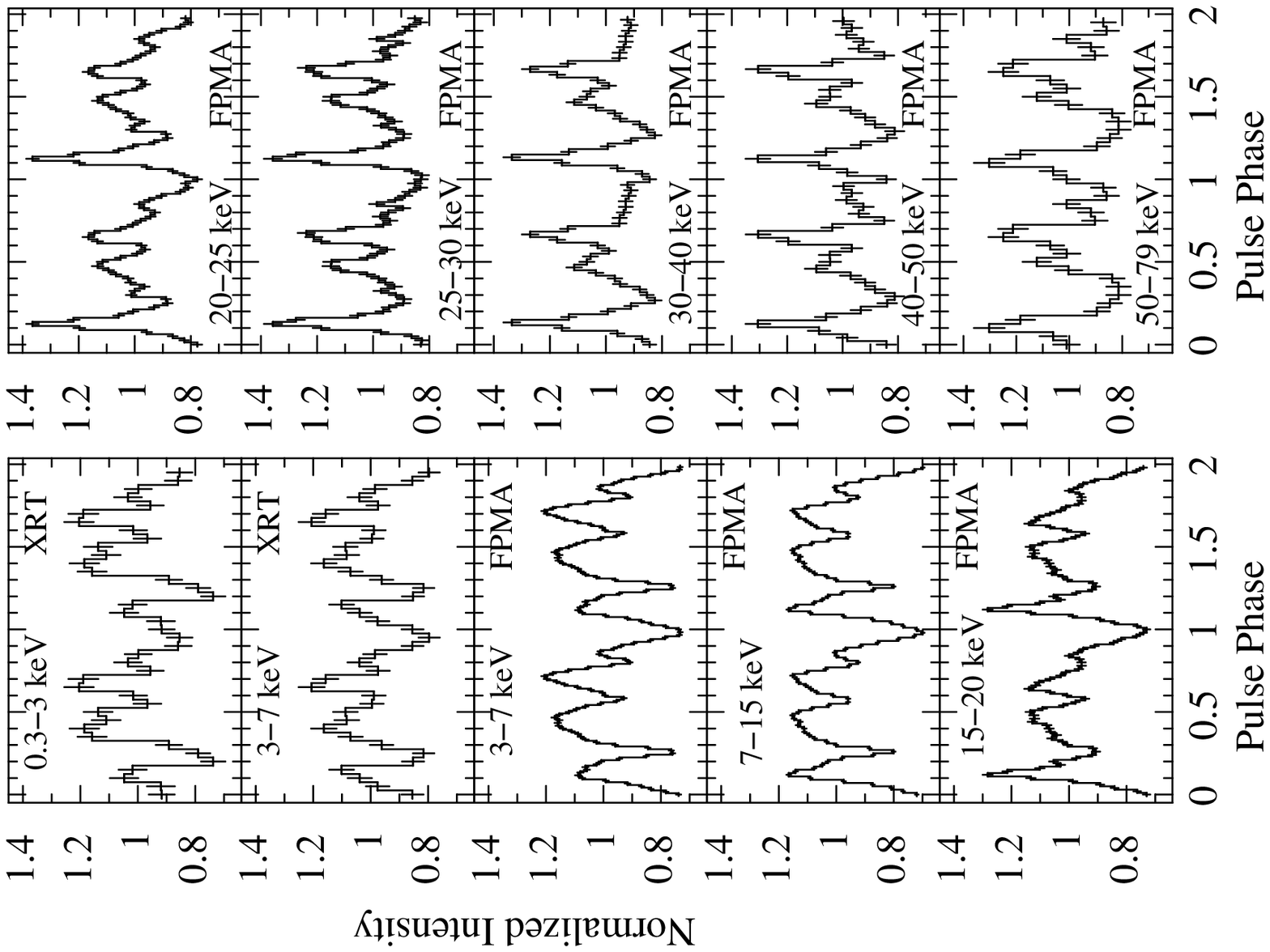}
\caption{Energy resolved pulse profiles of 2S~1417-624, obtained from {\it Swift} and {\it NuSTAR} observations of the pulsar during 2018 May outburst. A strong energy dependence could be traced easily among successive profiles. The error bars represent 1$\sigma$ uncertainties. Two pulses are shown in each panel for clarity.} 
\label{fig3}
\end{figure}
%-----------------------------------------------------------------------------------%

\subsection{{\it Swift} observation}

\pulsar was observed with {\it Swift}/X-Ray Telescope (XRT; \citealt{Burrows2005}) on the same day with an effective exposure of $\sim$1.8 ks. The XRT observes in 0.3--10~keV energy band using CCD detectors, with an energy resolution of 140~eV at 6~keV \citep{Burrows2005}. The CCD array consist of 600$\times$600 imaging pixels, each with the dimension of 40$\mu$m $\times$  40$\mu$m. {\it Swift}/XRT was operated in window timing mode during the observation of \pulsar. The unfiltered events were reprocessed by using {\tt XRTPIPELINE}. Standard procedures were then followed to extract the source products, as suggested by the instrument team\footnote{{http://www.swift.ac.uk/analysis/xrt/}}. The average source count rate during the entire observation was $\sim$13 counts~s$^{-1}$ (i.e. $<$100 counts~s$^{-1}$), which eliminates the possibility of photon pile-up on the CCD. Source products were accumulated by considering a rectangular region of 20$\times$40 pixels centered on the source position. Background was estimated in similar manner by considering the same size in the outer regions of the CCD. ARF files were accumulated by using {\tt XRTMKARF} task and the RMF files were sourced from the latest calibration data base files.

%-----------------------------------------------------------------------------------%
\begin{figure}
\centering
\includegraphics[height=3.2in, width=2.4in, angle=-90]{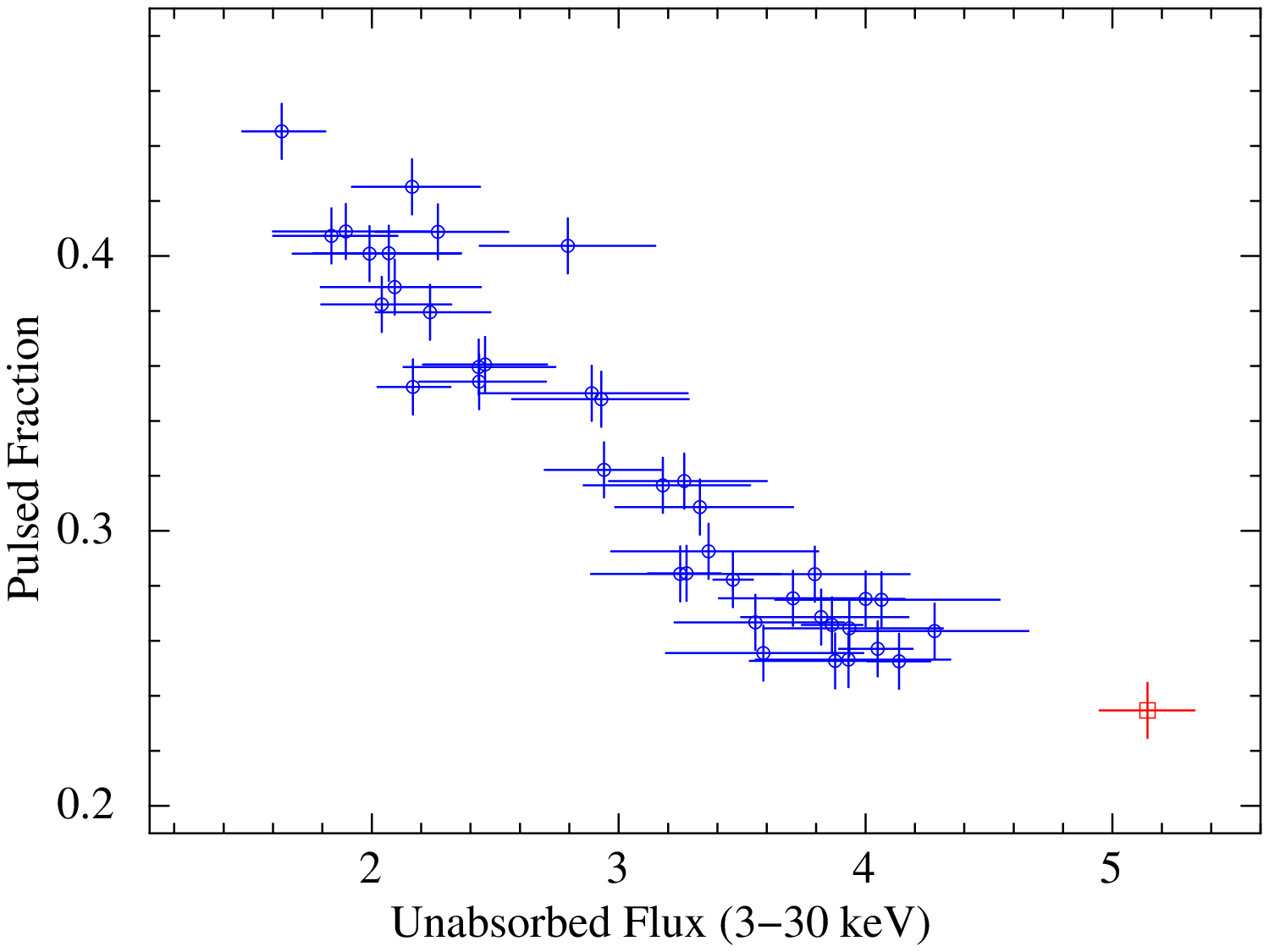}
\caption{The evolution of pulsed fraction with the 3--30~keV unabsorbed flux (in 10$^{-9}$~erg~cm$^{-2}$ s$^{-1}$ units). The blue circles corresponds to the pulsed fraction measured in 2-60 keV energy band with the {\it RXTE}/PCA during the giant outburst in 2009, while the red rectangle shows the measured value in 3-60 keV band via {\it NuSTAR} observation in 2018 outburst. } 
\label{fig4}
\end{figure}
%-----------------------------------------------------------------------------------%

\section {Timing analysis}

By following the above described procedures, source and background light curves with a time resolution of 0.1~s were extracted from the {\it NuSTAR} data. Barycentric correction was then applied on the background subtracted light curves. The $\chi^2$-maximization technique \citep{Leahy1987} was utilized to estimate the spin period of the pulsar. Pulsations at a period of 17.475(6)~s were detected in the source light curve. We also estimated the pulse period of the pulsar to be 17.47(5)~s, from the simultaneous {\it Swift}/XRT observation of the pulsar (see Section~2.2 for reduction procedures). The error on pulse period is estimated by fitting a Gaussian function on the chi-squared versus period distribution and is quoted for 1~$\sigma$ level. 

We have also estimated the spin period of the pulsar by using independent methods such as {\tt CLEAN} \citep{Roberts1987} and Lomb-Scargle periodogram \citep{Lomb1976, Scargle1982, Horne1986} as implemented in the {\tt PERIOD} program distributed with {\sc starlink} Software Collection\footnote{\url{http://starlink.eao.hawaii.edu/starlink}} \citep{Currie2014}. A consistent value of spin period of 17.475(1) was obtained from above methods. We then calculated the false Alarm Probability \citep{Horne1986} in order to check the significance of the power peak in the periodogram which was found to be above 95 per cent. Note that the uncertainty given by these methods corresponds to the minimum error on the period. It is also worth mentioning that the error on the estimated period is very difficult to calculate. Only reliable estimate could be from the simulation of large number of light curves via Monte Carlo or randomization methods (see, e.g.,  \citealt{Boldin2013}). Based on the agreement between estimated period from independent methods, we adopted 17.475(6)~s (larger value of error among all the methods) as the spin period of 2S~1417-624 in our study. Pulse profiles were then constructed by folding the light curves at this  period. 

In order to understand the evolution of pulse profiles with luminosity, an {\it RXTE} observation at the peak of 2009 giant outburst (see Section~2 and Table~\ref{log}) was also analyzed and presented in Figure~\ref{fig2}. The {\it NuSTAR} observation of the pulsar was performed at a higher flux level ($\approx$5.14$\times$ 10$^{-9}$~erg~cm$^{-2}$ s$^{-1}$ in 3--30~keV range) as compared to the {\it RXTE} observation ($\approx$4.13$\times$ 10$^{-9}$ erg~cm$^{-2}$ s$^{-1}$ in 3--30 keV range; see section~4 for detailed spectral analysis). Irrespective of the flux difference, well agreement between the two profiles was observed except for the appearance of an additional dip in 0.7-0.8 phase range of the pulse profile obtained from the {\it NuSTAR} observation (Figure~\ref{fig2}). This effectively produces a four-peaked pulse profile as opposed to the three-peaked profile observed with the {\it RXTE} during the peak of 2009 giant outburst. Although, \pulsar has shown single- to triple-peaked profiles during its earlier outbursts (\citealt{Finger1996, Inam2004, Raichur2010, Gupta2018}), pulse profiles with four peaks had never been observed before. This confirms the earlier results that the pulse profile of the pulsar is highly dependent on the source luminosity  (\citealt{Inam2004, Gupta2018}). This also motivated us to explore this intensity dependent evolution of pulse profiles further through phase-averaged and phase-resolved spectral studies (see Section~4). 

%-----------------------------------------------------------------------------------%
\begin{figure}
\centering
\includegraphics[height=3.2in, width=2.4in, angle=-90]{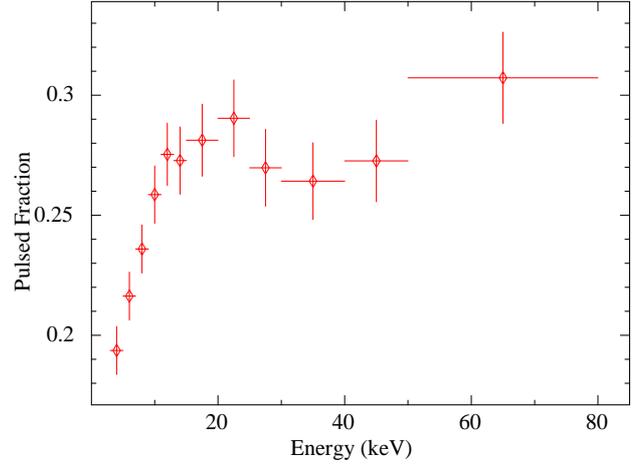}
\caption{Dependence of the pulsed fraction of \pulsar on energy during the {\it NuSTAR} observation in 2018. The error bars represent 1$\sigma$ uncertainties. } 
\label{fig5}
\end{figure}
%-----------------------------------------------------------------------------------%

We investigated the energy dependence of the pulse profiles of the pulsar by folding the energy resolved light curves obtained from the {\it NuSTAR} and {\it Swift} observations. This was done in order to understand the evolution of individual peaks and the emission geometry of the pulsar with energy. From Figure~\ref{fig3}, it is clear that the pulse profiles are strongly dependent upon the photon energy. Four peaks in the pulse profiles are visible up to $\approx$15~keV along with the change in relative intensities of the individual peaks. The evolution of first peak in 0.0--0.2 phase range is found to be different as compared to other peaks in the sense that it gets narrower and prominent with increasing energy.  The hard X-ray pulse profile (50--79~keV range) was found to be comparatively simple and appeared double peaked as the intensities of other peaks are reduced at higher energies. This is possibly due to the broadening of the minima (dip) after the first peak with increasing energy. The pulsations were detected in the light curves up to 79~keV. We notice a remarkable similarity in the evolution of the pulse profiles of the pulsar during the 2018 {\it NuSTAR} observation and the {\it RXTE} observations during 2009 outburst \citep{Gupta2018}, suggesting a similar accretion geometry during these outbursts. 

Furthermore, we estimated the pulse-fraction of the pulsar for the {\it NuSTAR} observation. Although, {\it NuSTAR} is operational in 3--79~keV energy range, we have estimated the value of pulse-fraction in 3--60~keV energy band. This was done, in order to have a direct comparison with the previous outburst studies in 2009 \citep{Gupta2018}. The pulse-fraction is defined as the ratio between the difference of maximum and minimum intensities to the sum of maximum and minimum intensities in the folded light curve. The pulse-fraction in 3--60~keV range for the {\it NuSTAR} observation was estimated to be 23.4$\pm$0.4 per cent and is shown in Figure~\ref{fig4}. The estimated value of pulsed fraction seems to follow the negative correlation trend with the pulsar luminosity as observed during the previous 2009 outburst. Apart from the dependence of the pulsed fraction on luminosity, we have also studied the change in pulsed fraction with energy (Figure~\ref{fig5}). It is evident  that the pulsed fraction of \pulsar increases with energy, similar to other X-ray pulsars \citep{Lutovinov2009}. 

Energy dependence of the pulsed fraction has been studied extensively for a large sample of transient Be/X-ray binary pulsars by \citet{Lutovinov2009}. Most of the pulsars were found to exhibit a monotonic increase in the pulsed fraction with energy. While some of them additionally showed local features near the cyclotron line energy and its harmonics. The presence of these features was attributed to the effect of resonance absorption. This qualitative explanation for the increase in pulsed fraction with energy, is based on a simple geometrical model for the neutron stars with dipole magnetic fields (see \citet{Lutovinov2009} and references therein).

\section {Spectral Analysis} 

\subsection {Phase-averaged spectroscopy of 2S~1417-624}

We performed phase-averaged spectroscopy of \pulsar in 0.9--79~keV energy range using simultaneous observations of the pulsar with {\it NuSTAR} and {\it Swift}/XRT, at the peak of the 2018 May giant outburst. The source and background spectra were accumulated by following the standard procedures described in Section~2. Using appropriate background, response matrices and effective area files, spectral fitting was carried out by {\tt XSPEC} version 12.8.2 package. For simultaneous spectral fitting, all model parameters were tied together for FPMA, FPMB and XRT spectra while the relative instrument normalizations were kept free. The cross-normalization constants between the detectors were consistent with the value suggested by the instrument team.  While fitting the {\it Swift}/XRT and {\it NuSTAR} data, we noticed a marginal mismatch between the distribution of data points below 3.5 keV. Therefore, for simultaneous spectral fitting, we did not used {\it NuSTAR} data below 3.5 keV. This mismatch have also been previously reported by several authors in other studies (\citealt{Kuhnel2017, Bellm2014}).

\begin{figure}
\centering
\includegraphics[height=3.4in, angle=-90]{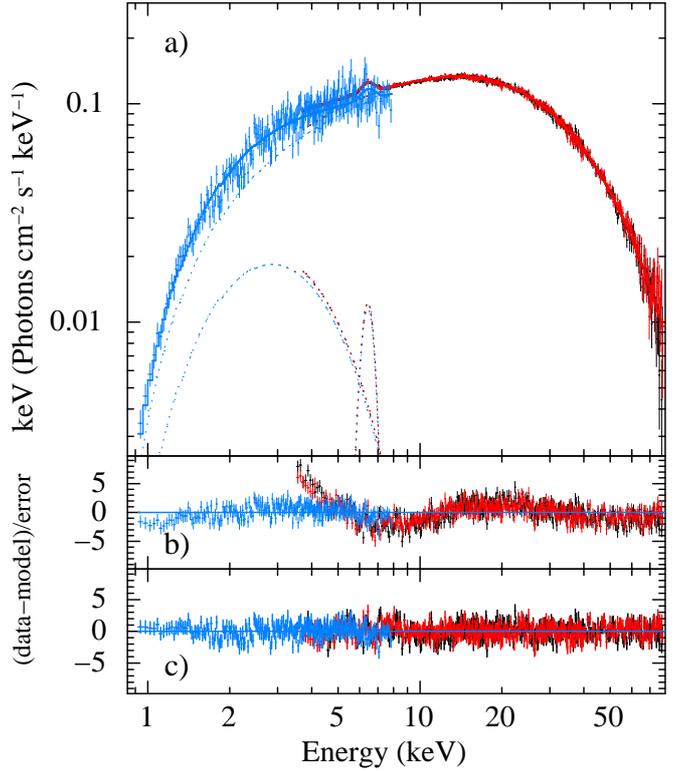}
\caption{ (a) : Broad-band continuum spectrum of 2S~1417-624 in 0.9--79~keV energy range, obtained from {\it Swift}/XRT (blue), {\it NuSTAR}-FPMA (red) and -FPMB (black) detectors. The top panel shows the source spectra obtained from above mentioned detectors along with the best fitted model comprising of a cutoff power law continuum model and a blackbody component (Cutoff+BB) modified with the interstellar absorption and a Gaussian function for the iron emission line at 6.4~keV. Panels (b) and (c) shows the spectral residuals obtained after fitting the broad-band spectrum, without and with the inclusion of blackbody component, respectively.} 
\label{fig6}
\end{figure}

The broad-band continuum spectrum of \pulsar has been described with a power law model modified with a cutoff at higher energies. Therefore, we initially approximated the spectrum with a cutoff power law model. Additionally, we detected a fluorescent iron emission line at 6.4~keV. The result of this fitting i.e, phabs$\times$(CutoffPL $+$ ga), yielded a reduced $\chi^2$ value of $>$2 and the corresponding residuals are shown in Figure~\ref{fig6}(b). The presence of wave-like residuals in 1--15~keV energy range of the spectrum and high value of reduced $\chi^2$ made the fitting unacceptable. In order to account for these residuals, we included a thermal blackbody (BB) component with a temperature of $\sim$1~keV. Addition of this BB component improved the fitting with corresponding reduced $\chi^2$ value of $\approx$1.1 (Figure~\ref{fig6}(c)). We have also attempted to use partial covering (PC) absorber instead of the BB component, with a comparable statistical significance. Though the pulsar was observed at the highest luminosity till date (at the peak of 2018 giant X-ray outburst), there was no signature of presence of any cyclotron resonance scattering feature in the pulsar spectrum in 0.9--79 keV range. 

We investigated several other standard continuum models such as high energy cutoff power law (HECut; \citealt{White1983}), a HECut power law with a smoothing Gaussian at cutoff energy (NewHcut; \citealt{Burderi2000, Jaisawal2015}) and negative and positive exponential cutoff (NPEX) models, in order find a suitable continuum for \pulsar. For all considered models, the addition of BB or a PC component and an iron line at $\sim$6.4~keV was necessary for the spectral fitting. The spectral parameters obtained from fitting the broad-band spectrum of the pulsar with these models are listed in Table~\ref{spec-par}. The value of equivalent hydrogen column density obtained from fitting various spectral models, was found to be in range $\approx$0.6--0.9$\times$10$^{22}$ cm$^{-2}$ units, which is lower than the  value of Galactic absorption in the source direction ($\approx$1.4$\times$10$^{22}$ cm$^{-2}$; \citealt{Willingale2013}). This indicated that the presence of an additional absorber close to the neutron star is unlikely. However, the spectral fitting with PC model leads to significantly higher additional column density of $\approx$168$\times$10$^{22}$ atoms cm$^{-2}$ units. Therefore, we consider the two-component model, i.e, CutoffPL and a thermal blackbody (CutoffPL$+$BB) to be the best-fit continuum model. Panel (a) of Figure~\ref{fig6} shows the spectral fitting of \pulsar by CutoffPL$+$BB model. The residuals obtained by fitting the model without and with the contribution of blackbody component, are shown in panels (b) and (c) of  Figure~\ref{fig6}, respectively.

%-----------------------------------------------------------------------------------%
%-----------------------------------------------------------------------------------%

\subsection {Broad-band continuum modelling with physical models}

The formation of accretion column above the magnetic poles of the accreting neutron star, is responsible to shape the characteristic spectrum of the X-ray pulsars \citep{Becker2007}. The detection of the thermal blackbody component in the phase-averaged spectrum of \pulsar motivated us to explore the properties of accretion column via physically motivated models such as $BWmod$ \citep{Becker2007} and $COMPMAG$ model \citep{Farinelli2012}. 

$BWmod$ solves the radiative transfer equation (RTE) inside the accretion column analytically and it uses a specific velocity profile that is linearly dependent on optical depth of the column \citep{Becker2007}. It has been implemented on the bright pulsars such as 4U~0115$+$63, Her~X-1 and EXO~2030$+$375 (\citealt{Ferrigno2009, Wolff2016, Epili2017}). Due to large uncertainty in the distance estimates for \pulsar, we tried to use $BWmod$ by considering two different distance values i.e., 5 and 11~kpc. However, our attempts to use $BWmod$ to fit the pulsar spectrum failed with the reduced $\chi^2$ value obtained from the fitting as $>$2 in each case.  

Afterwards, we attempted to implement the $COMPMAG$ model to the pulsar spectrum. $COMPMAG$ model solves the RTE numerically and it offers different velocity profiles, characterized by an index $\eta$ and a terminal velocity $\beta$. The other free parameters of the model are (i) temperature of seed blackbody spectrum $kT_{bb}$, (ii) electron plasma temperature $kT_e$ and (iii) vertical optical depth of the accretion column $\tau$. While fitting this model, we assumed the velocity profile of the accretion column to be linearly dependent on the optical depth ($\beta(\tau)$$\propto$$\tau$, i.e. beta flag 2; \citealt{Farinelli2012}). This gives rise to an acceptable fit with reasonable values of spectral parameters (listed in Table~\ref{spec-par}). However, we also remark that this model is better suited for lower luminosity observations where the accretion flow is expected to get halted by Coulomb interactions, as opposed to a radiation-dominated shock expected in case of higher luminosities.

%--------------------------------------------------------------------------------------------%
\begin{table*}
\centering
\caption{Best-fitting spectral parameters (90\% errors) obtained from the simultaneous {\it NuSTAR} and {\it Swift}/XRT observations of 2S~1417-~624. The fitted models are (i) High-energy cutoff power law with a blackbody component, (ii) cutoff power law model with blackbody, (iii) NEWHCUT model with a blackbody component, (iv) Cutoff power law modified with a partial covering absorption, and (v) COMPMAG model along with photoelectric absorption component and a Gaussian component for iron emission line.}

\begin{tabular}{ |l | ccccc}
\hline 
Parameters                      &  \multicolumn{5}{c}{Spectral Models}    \\ 
&  \\\cline{2-6}  \\
                                &HECut+BB     &Cutoff+BB	&NEWHCut+BB   &CutoffPL(with PC)    &COMPMAG	 \\
\hline
%\\
N$_{{H}_{1}}$$^a$       	         	&0.63$\pm$0.05         &0.72$\pm$0.04	&0.69$\pm$0.04  &0.83$\pm$0.04         &0.67$\pm$0.06  \\
N$_{{H}_{2}}$$^b$		&--			&--		&--		&168.3$\pm$12.4			&--  \\
Cov. Fraction			&--			&--		&--		&0.16$\pm$0.01			&--	\\
Photon index         		 &0.12$\pm$0.01       &0.12$\pm$0.01 	&0.10$\pm$0.01  &0.37$\pm$0.01           &--		\\
E$_{cut}$ (keV)	                &3.1$\pm$0.5         &15.3$\pm$0.2 	&5.4$\pm$0.4   	 &17.04$\pm$0.15         &--   \\
E$_{fold}$ (keV)	      &15.3$\pm$0.1              &-- 		&15.2$\pm$0.2		&--	         &--  \\
BB temp. (keV)       		&0.94$\pm$0.04	    &0.96$\pm$0.04   	&0.94$\pm$0.03		&--	     &--   \\        
\\
COMPMAG kT$_{bb}$ (keV)          &--                  &--        &-- 	&--              &0.86$\pm$0.06  \\
COMPMAG kT$_{e}$ (keV)  	 &--                  &--        &--  	&--    		 &4.45$\pm$0.12  \\
COMPMAG $\tau$                   &--                  &--        &--  	&--          &0.90$\pm$0.02    \\
Column radius (km)               &--                  &--     	 &-- 	&--	      &2.18$\pm$0.05    \\
\\
{\it Iron line parameters} \\
Line energy (keV)              &6.41$\pm$0.02       &6.41$\pm$0.02 	&6.40$\pm$0.02   &6.39$\pm$0.02   	&6.41$\pm$0.02 \\
Eq. width  (eV)               &82$\pm$9	    &82$\pm$12  	&76$\pm$8  &69$\pm$15       &79$\pm$7 \\
\\
{\it Component Flux} (1-79 keV)$^c$	
\\
Power law flux	        &8.11$\pm$0.02	    &8.11$\pm$0.01	&8.09$\pm$0.03		&8.24$\pm$0.03		&--	\\
Blackbody flux		&0.13$\pm$0.01    &0.13$\pm$0.01	&0.15$\pm$0.02 	&--		&--	\\
\\
{\it Source flux}$^c$
 \\
Flux (1-10 keV)   		&1.46$\pm$0.02      &1.47$\pm$0.01  	&1.47$\pm$0.01  &1.46$\pm$0.01       &1.47$\pm$0.02		 \\
Flux (10-79 keV)  		&6.78$\pm$0.01      &6.77$\pm$0.02  	&6.78$\pm$0.01  &6.79$\pm$0.01       &6.80$\pm$0.02		 \\ 
\\
Reduced $\chi^2$ (\it d.o.f)         &1.10 (1065)          &1.10 (1066)  	&1.09 (1065) 	&1.11 (1064)          &1.11 (1066)  \\
%\\
\hline
\end{tabular}
\flushleft
$^a$ : Equivalent hydrogen column density in the source direction(in 10$^{22}$ atoms cm$^{-2}$ unit) \\
$^b$ : Additional hydrogen column density (in 10$^{22}$ atoms cm$^{-2}$ unit) \\
$^c$ : Unabsorbed flux in unit of 10$^{-9}$ ~erg~cm$^{-2}$~s$^{-1}$. 
\label{spec-par}
\end{table*}
%-------------------------------------------------------------------------------%    

\subsection {Pulse phase-resolved Spectroscopy}

\begin{figure}
\centering
\includegraphics[height=3.2in, angle=-90]{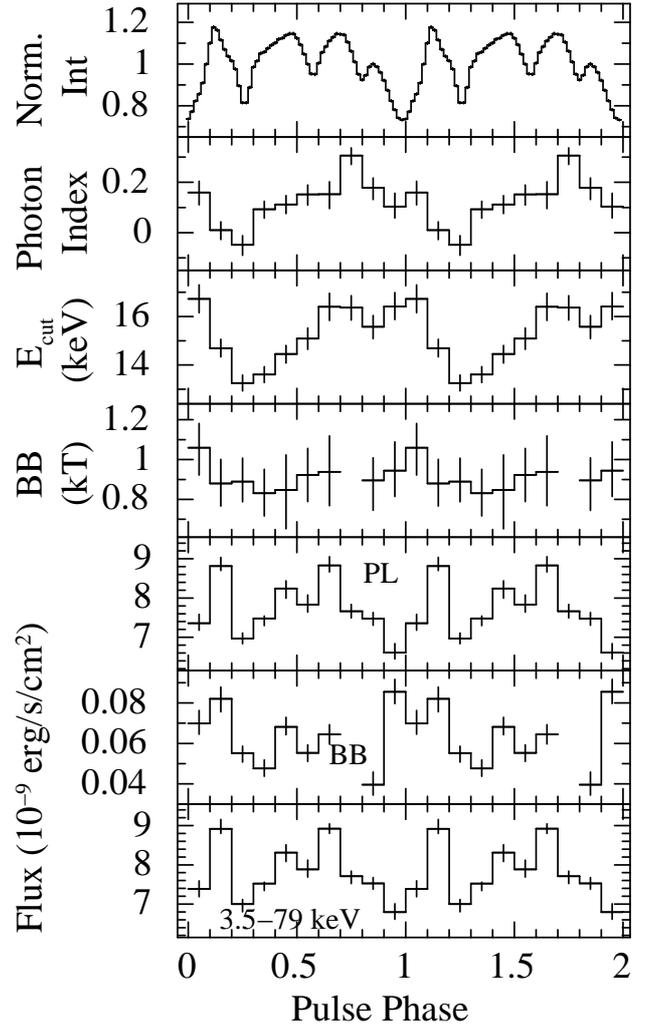}
\caption{Spectral parameters obtained from the phase-resolved spectroscopy of 
2S~1417-624 during {\it NuSTAR} observation in 2018 May by using cutoff+BB continuum model. Top panel shows the pulse profile of the pulsar in 3--79~keV energy range. The values of power-law photon index, cutoff energy ($E_{cut}$) and blackbody temperature are shown in the second, third and fourth panels from top, respectively. The power-law flux (PL), blackbody component flux (BB) and total flux in 3.5--79~keV range are presented in fifth, sixth and seventh panels, respectively. All fluxes are quoted in the units of 10$^{-9}$ erg cm$^{-2}$ s$^{-1}$. The errors in the spectral parameters are estimated for 90 per cent confidence level.} 
\label{fig7}
\end{figure}

In order to investigate the cause of four peaks in the pulse profiles, the nature of thermal blackbody component and the variation of other spectral parameters during the 2018 May outburst, we carried out pulse phase-resolved spectroscopy of \pulsar by using {\it NuSTAR} observation. For this, the source spectra were accumulated in 10 phase-bins by applying phase filters on the barycentric corrected event file in {\tt xselect} package. For phase-resolved spectral fitting, we used the same energy range as chosen for the phase averaged spectroscopy of the {\it NuSTAR} data. Each individual phase-sliced spectra were fitted by using appropriate background, response and effective area files. As in case of phase averaged spectral fitting, both the cutoff+BB and HECut+BB models along with the photoelectric absorption component and a Gaussian function at 6.4~keV were used to fit the spectra of each phase bins. It was found that both the models fit all the phase-resolved spectra well, yielding comparable values of fitted parameters. While fitting, the value of equivalent hydrogen column density was fixed to the phase averaged value (N$_{{H}_{1}}$; Table~\ref{spec-par}). The phase-resolved spectral parameters obtained from the fitting are shown in Figure~\ref{fig7} for cutoff+BB model, along with the pulse profile of the pulsar at the top panel. 

Although the shape of the pulse profile of 2S~1417-624 during {\it NuSTAR} observation was significantly different from those during previous outbursts in {\it RXTE} era, the variation in the spectral parameters over the pulse phases were marginal. From Figure~\ref{fig7}, it can be seen that the values of power law photon index and cutoff energy were found to be variable in range $-$0.1--0.3 and 13--17~keV, respectively. The blackbody component was detectable in all phase bins except for one (fourth panel of Figure~\ref{fig7}), which indicates that the blackbody flux was either too low or not significant beyond 3 keV. While the blackbody temperature was found to be almost constant (within errors) throughout the pulse phases. The power law and model fluxes obtained in 3.5--79~keV energy range, from the fitting of individual phase bins, follow the shape of pulse profiles (fifth and seventh panel of Figure~\ref{fig7}, respectively). Apart from the shape of total flux profiles over the pulse phase, an enhancement in the blackbody (BB) flux component could be clearly seen in 0.1--0.2, 0.4--0.5 and 0.9--1.0 phase ranges (sixth panel; Figure~\ref{fig7}). This enhanced value of BB flux can be associated with the peculiar evolution of peaks in the energy resolved pulse profiles (Figure~\ref{fig3}). All the flux values quoted in the paper are calculated by using the {\tt cflux}  convolution model. Other parameters such as iron line energy and its equivalent width were found to be consistent with the phase-averaged values and do not show any significant variation over the pulse phase.

%-----------------------------------------------------------------------------------%
%-----------------------------------------------------------------------------------%

\section {Discussion and Conclusions}

In the present work, we have performed timing and spectral analysis of \pulsar during its 2018 giant outburst by using simultaneous observations with the {\it NuSTAR} and {\it Swift} observatories. These observations were performed at a flux level of approximately 350~mCrab (i.e. at the peak of the giant outburst) as recorded in {\it Swift}/BAT \citep{Nakajima2018}. It is one of the brightest giant X-ray outburst observed from the source till date. One of the most interesting aspect of the present study is the featuring of four-peaks in the pulse profile of the pulsar which have never been observed before, during any type of outburst from 2S~1417-624. Our results are in agreement with the previous findings that the pulse-profiles of the pulsar are strongly luminosity dependent \citep{Gupta2018}. These pulse profiles were found to be strongly dependent upon photon energy as well (see Figure~\ref{fig3}). 

The transient Be/X-ray binary pulsars are known to show multiple dips and peaks in their pulse profiles during X-ray outbursts. These peaks are known to be strongly energy dependent in nature. A few of such pulsars showing energy and luminosity dependence of pulse profiles are EXO~2030+375 (\citealt{Naik2013, Epili2017}), GX~304-1 \citep{Jaisawal2016}, A0535+262 \citep{Naik2008}, 1A~1118-61 \citep{Maitra2012}. Pulse phase-resolved spectroscopy of observations of these pulsars during outbursts revealed the presence of additional matter at the dip phases in the pulse profiles. This suggest that the dips or dip-like features in pulse profiles are due to the absorption/obscuration of X-ray photons by additional local matter present around the neutron stars. Following this, we modelled the phase-resolved spectra of \pulsar with a partial covering cutoff power law model. As described in section 4.3, we fixed the value of equivalent hydrogen column density (N$_{{H}_{1}}$) to the phase averaged value and left additional hydrogen column density (N$_{{H}_{2}}$) free to vary. We obtained an exceptionally high value of additional column density during the primary dip (0.95-1.05 pulse phase range: top panel of Figure~\ref{fig7}). During other phases, there was no significant variation in the additional column density to draw any conclusion on this regard. Future studies with sensitive soft X-ray instruments like {\it NICER} could provide conclusive results on the matter distribution around the poles of the neutron star.

\pulsar is a unique pulsar in the sense that it shows highly luminosity dependent pulse-profiles, ranging from a single broad peak at low luminosities to multiple (four) peaks at higher luminosities (\citealt{Gupta2018} and present work). A detailed pulse profile modelling is required to study this diverse behavior of the source. However, such a study requires a physically motivated self-consistent working model to explain the pulsar beam geometry which takes into account of the beam pattern and gravitational light bending effects, which is extremely complex and outside the scope of this paper. 

Despite the complexity of pulsar emission mechanism, the broad-band spectrum of \pulsar has been successfully described with the standard continuum models such as cutoff power law and high energy cutoff power law models (\citealt{Inam2004, Gupta2018} and references therein). However, in the present study, we found that these models are not suitable to describe the spectrum obtained at the peak of the 2018 giant X-ray outburst. Rather, it was found that a composite model consisting of a cutoff power law and blackbody component was required to describe the spectrum well. The {\it NuSTAR} observation presented here was carried out when the pulsar was brighter compared to the {\it RXTE} observations during the 2009 giant X-ray outburst. The additional blackbody component is, therefore, very likely due to the enhanced luminosity of the pulsar though the sensitivities of {\it NuSTAR} detectors at soft X-ray ranges are better than the {\it RXTE}/PCA.

Several of the accretion powered X-ray binary pulsars (XBPs) have been observed to show an excess in the soft X-ray ranges of their spectrum. This is known as soft excess and has been modeled with an additional thermal component in the pulsar spectrum. This soft excess is thought to be a very common intrinsic feature of XBPs. However, its detectability depends upon the source flux and column density (N$_H$) in the source direction (see \citealt{Hickox2004} for a review). The possible mechanism behind the origin of this component could be : (i) emission from accretion column, (ii) thermal emission from collisionally energized diffuse gas around the neutron star, (iii) reprocessing of hard X-rays by diffuse cloud, and (iv) reprocessing of hard X-rays in optically thick accretion disk \citep{Hickox2004}. By using the value of blackbody normalization obtained from the phase-averaged spectral fitting and considering the source distance of 5~kpc, the radius of blackbody emitting region is estimated to be $\approx$2~km. In our study, the presence of hard Comptonized spectrum as well as an emitting radius of $\approx$2~km makes difficult to consider accretion column as a primary source of soft excess emission. The non-pulsating nature of blackbody component also suggests the same. Alternatively, it is possible that the neutron star surface is contributing for reflection of X-ray photons from the column. This may explain the presence of non-pulsating thermal emission from the pulsar. A similar analogy is suggested by \citet{Poutanen2013} on cyclotron line scattering feature.

In the present study, we have witnessed a change in spectral continuum at higher luminosity. By considering a distance of 5~kpc and 11~kpc, the source luminosity in 1--79 keV range was estimated to be $\approx$2.46$\times$10$^{37}$ erg~s$^{-1}$ and 1.19$\times$10$^{38}$ erg~s$^{-1}$, respectively. This luminosity ($\sim$10$^{37}$ erg~s$^{-1}$) is crucial in understanding the accretion state transitions in accretion powered X-ray pulsars and is known as critical luminosity \citep{Becker2012}. At such high luminosities, the radiation dominated shocks may form near the neutron star surface, giving rise to dramatical changes in pulsar beam configuration and spectral parameters (see \citealt{Reig2013, Epili2017}). Thus, the spectral changes seen in the present study could be associated with the critical luminosity of the source. Similar results were found during the previous giant outburst in 2009 when the pulsed fraction of the pulsar was anti-correlated with the source flux in a luminous regime \citep{Gupta2018}.

It is also important to note that a positive correlation was detected between pulsed fraction and flux below 1.2$\times$10$^{-9}$~erg~cm$^{-2}$~s$^{-1}$ (see, Figure~2 of \citealt{Inam2004}). Above this limit, we detected a clear anti-correlation in the present study. The observed anti-correlation can be  attributed to the increase in unpulsed component from the pulsar in the form of fan beam emission close to the critical luminosity. The timing analysis, in the present study  (Figure~\ref{fig4}) also supports this idea in context of earlier finding by \citet{Gupta2018}.

In summary, we have presented a detailed spectral and timing analysis of 2S~1417-624 during the 2018 giant outburst by using simultaneous observations from {\it NuSTAR} and {\it Swift} observatories. The pulse profiles were found to four-peaked and strongly dependent on energy. The broadband energy spectrum of the pulsar is well described with a composite spectrum consisting of cutoff power law and a thermal blackbody component along with an iron fluorescence line at $\sim$6.4~keV. Considering the source distance to be $\sim$5 kpc, we estimated the radius of the blackbody emitting region to be $\approx$2 km. The reflection of X-ray photons from the stellar surface possibly contributes to the observed soft excess. Based on the results obtained from spectral fitting via physical and empirical models, we interpret that the source is consistently accreting close to the critical luminosity regimes.

\section*{Acknowledgments}
We thank the anonymous referee for his/her useful suggestions that improved 
the quality of the paper. The research work at Physical Research Laboratory 
is funded by the Department of Space, Government of India. This research has 
made use of data obtained through HEASARC Online Service, provided by the 
NASA/GSFC, in support of NASA High Energy Astrophysics Programs.

\def\apj{ApJ} \def\mnras{MNRAS}
\def\aap{A\&A} \def\apjl{ApJ} \def\aj{aj} \def\physrep{PhR}
\def\pre{PhRvE} \def\apjs{ApJS} \def\pasa{PASA} \def\pasj{PASJ}
\def\nat{Nat} \def\ssr{SSRv} \def\aapr{AAPR} \def\araa{ARAA}

\end{document}